\documentclass{IEEEcsmag}

\usepackage[colorlinks,urlcolor=blue,linkcolor=blue,citecolor=blue]{hyperref}
\expandafter\def\expandafter\UrlBreaks\expandafter{\UrlBreaks\do\/\do\*\do\-\do\~\do\'\do\"\do\-}
\usepackage{upmath,color}

\jvol{XX}
\jnum{XX}
\paper{8}
\jmonth{March}
\jname{IEEE Security \& Privacy}
\jtitle{LLM-Powered Agent-based Framework for Misinformation and Disinformation Research: Opportunities and Open Challenges}
\pubyear{2024}

\usepackage{etoolbox}
\usepackage{soul} 

\newtoggle{finalPaper}

\setstcolor{blue}

\toggletrue{finalPaper} 

\iftoggle{finalPaper} {
	\newcommand{\addtxt}[1]{#1}
	\newcommand{\change}[2]{#2}
	\newcommand{\rmvtxt}[1]{}}
{
    \soulregister\cite7 
    \soulregister\ref7
    
	\newcommand{\addtxt}[1]{\textcolor{blue}{#1}}
    \newcommand{\change}[2]{\texorpdfstring{\st{#1}\textcolor{blue}{#2}}{#2}}
    \newcommand{\rmvtxt}[1]{\texorpdfstring{\st{#1}}{#1}}
}

\begin{document}

\sptitle{Theme Article: Synthetic Realities and Artificial Intelligence-Generated Contents}

\title{LLM-Powered Agent-based Framework for Misinformation and Disinformation Research: Opportunities and Open Challenges}

\author{Javier Pastor-Galindo}
\affil{Dept. of Information and Communications Engineering, University of Murcia, 30100, Spain}

\author{Pantaleone Nespoli}
\affil{Dept. of Information and Communications Engineering, University of Murcia, 30100, Spain}

\author{José A. Ruipérez-Valiente}
\affil{Dept. of Information and Communications Engineering, University of Murcia, 30100, Spain}

\markboth{Synthetic Realities and Artificial Intelligence-Generated Contents}{Synthetic Realities and Artificial Intelligence-Generated Contents}

\begin{abstract}
This article presents the affordances that Generative Artificial Intelligence can have in misinformation and disinformation contexts, major threats to our digitalized society. We present a research framework to generate customized agent-based social networks for disinformation simulations that would enable understanding and evaluating the phenomena whilst discussing open challenges.
\end{abstract}


\maketitle


\section{GENAI: DECODING MIS/DISINFORMATION}

The advent of Generative Artificial Intelligence (GenAI) has fundamentally reshaped the field of digital content creation, impacting how we can produce images, videos, audio, and text. Presently, AI models can craft remarkably realistic content that aligns with the context provided in simple language prompts. Standout Large Language Models (LLMs) like GPT-4 (OpenAI), Claude (Anthropic), PaLM and LaMDA (Google), LLaMA (Meta AI), Chinchilla (Deep Mind), and Alpaca (Stanford), have greatly enhanced the generation of text that aligns with the given context. Similarly, image generation models such as DALLE 2 (OpenAI), Stable Diffusion (Runway), and IMAGEN (Google) have introduced a new approach for creating images that accurately depict real-life scenarios. Notably, text-to-video models like Phenaki (Google) and Gen-2 (Runway) have also demonstrated significant progress~\cite{gozalobrizuela2023chatgpt}.

The introduction of these generative technologies, equipped with open-source models and accessible interfaces, has positively influenced productivity across a range of areas like programming, entertainment, education, and arts. 
In academia and research, particularly for social scientists, these tools offer novel opportunities for creating realistic content, simulating human behavior, or tailoring behavioral experiments~\cite{Ziems2023}. Recent trials conducted by major corporations and universities have highlighted the potential of these AI tools in areas like self-guided life simulations, open-world experiments, psychological studies, and social simulations~\cite{Park2023}. 

In this context, it is easy to argue that GenAI, particularly LLMs, represent a promising technology against one of the major threats happening within social media nowadays, i.e., disinformation. In our digitalized society, the pervasive threat of disinformation is underscored by its multifaceted impact on democratic institutions, public trust, and global stability~\cite{pastor2020}. The urgency to address this issue is evident, particularly as highlighted in the World Economic Forum's Global Risks Report 2024\footnote{https://www.weforum.org/publications/global-risks-report-2024/}, where misinformation and disinformation consistently rank as the foremost severe short-term threat over the next two years. While consensus exists on the need for measures to combat disinformation, its impact remains challenging to analyze comprehensively. Responsible cooperation among stakeholders is crucial, and initiatives at both national and European levels aim to strengthen democracy against disinformation. Particularly, the integration of new developments in AI is recognized as a potential turning point, offering opportunities for enhanced detection and mitigation amid intensified challenges and risks.

Throughout this study, we delve into the potential of LLMs as an innovative method for comprehending, simulating, and evaluating disinformation within controlled experimental settings~\cite{ghaffarzadegan2023generative}. In a traditional context, disinformation has predominantly centered around the theoretical modeling of fake news propagation and influence, as well as leveraging social media data for detection and assessment. This field grapples with several issues including the complexity of scrutinizing incidents where there is no truth baseline to affirm the objectives, tactics, and actors involved in influence campaigns, the lack of labeled datasets for various manipulation efforts, the infeasibility of testing technical countermeasures in third-party platforms, or the necessity of human involvement to measure the cognitive impact of deceptive activities~\cite{sumers2023cognitive}. 

Conversely, LLMs are being used to realistically rule systems with agents embodying human behaviors, replacing mathematical models and static experiments~\cite{gao2023s3}. This advancement opens the door to creating any information environment controlling the context, users and functioning of message exchange, leading to generative agent-based social networks as sandboxes. In these controlled scenarios, red agents can be programmed to simulate custom disinformation attacks for further analysis of their evolution and influence on the LLM-driven network of individuals. Therefore, we posit that LLMs can potentially alleviate some of the technical obstacles for developing research to address the following tasks:

\begin{itemize}
    \item Model and simulate a realistic social network with users, interactions and information flows.
    \item Model, simulate and assess different types of misinformation cases and disinformation attacks against a realistic social network.
    \item Model, simulate and evaluate different types of technical countermeasures to misinformation cases and disinformation attacks.
\end{itemize}

This article proposes a conceptual LLM-powered framework for misinformation and disinformation research, delving into an extensive examination of research opportunities and identifies prominent unresolved challenges to respond to the aforementioned research questions. For simplicity, we will use only the term disinformation throughout the rest of this article.

\section{LLM-POWERED AGENT-BASED FRAMEWORK FOR MIS/DISINFORMATION RESEARCH}

\begin{figure*}[t!]
    \centering
    \includegraphics[width=\textwidth]{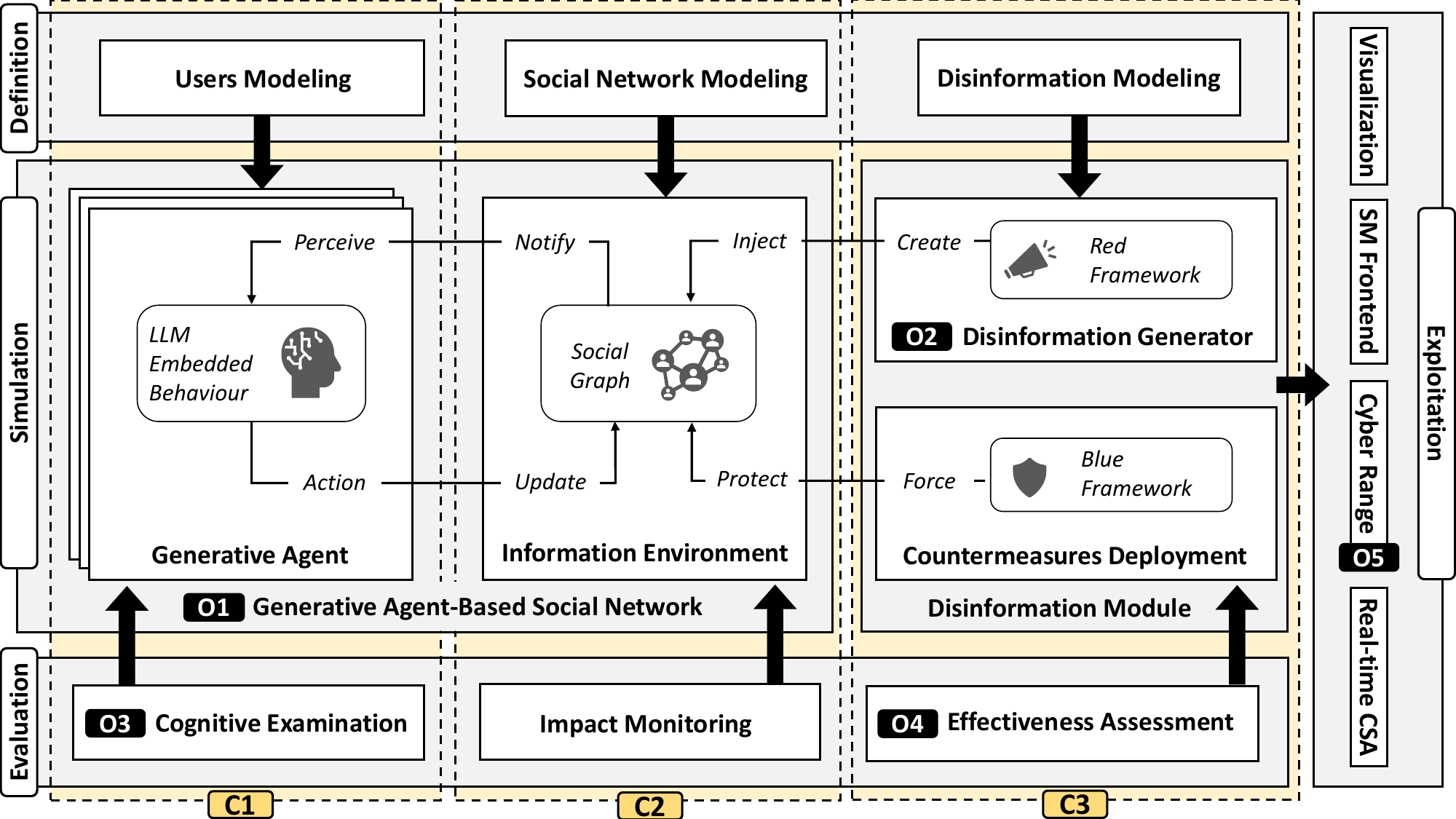}
    \caption{Layers, elements, opportunities and challenges of the LLM-Powered Agent-based Framework for Disinformation Research}.
    \label{fig:framework}
\end{figure*}

As mentioned before, the main proposal of this research is a conceptual framework for disinformation research. In particular, such framework represented in \figurename~\ref{fig:framework} is composed of five interconnected layers, each one exposing certain characteristics and functionalities. In the following, each of them is detailed to clarify the design's decisions.

\subsection{Definition layer}
First, the Definition layer is responsible for modeling the entities that compose the framework, which are then recreated in the simulation environment. Specifically, three entities are modeled, i.e., users, social network, and disinformation. Such a choice relies on the fact that one can consider those entities as the fundamental blocks to correctly build a full-fledged framework to boost disinformation research.

\subsection{Simulation layer}
Next, the Simulation layer contains the simulated entities, i.e., the LLM-powered agents, the social network itself, and the disinformation module. This layer receives the models created by the Definition layer to perform the simulation. The core component is the social graph. Obviously, the agents \addtxt{driven by a LLM} represent the participants of such a graph, perceiving a certain dynamic and acting to update the graph. Those interactions create a continuous loop that, in fact, recreates the inherent dynamics of a real social network based on the models' parameters. Moreover, the disinformation module is strongly connected with the social graph, too. This module includes both the offensive and defensive framework. In fact, it is responsible for the disinformation generation and countermeasures enforcement.

\subsection{Evaluation layer}
Then, the Evaluation layer is in charge of assessing the overall situation within the simulation environment from different perspectives. Concretely, the framework envisions a cognitive, impact, and effectiveness perspective. That is, the cognitive examination helps assess whether the LLM-powered agents follow a pre-configured cognitive theory (e.g., confirmation, availability, etc.) while generating social content. Also, the impact monitoring component is in charge of evaluating the impact of the agents' behaviour within the social graph. Lastly, the effectiveness assessment is responsible for the remediations enforcement, i.e., actions executed to counteract disinformation dynamics.

\subsection{Exploitation layer}
Last but not least, the Exploitation module connects the framework with other valuable tools to fully leverage its potentialities and involve human actors from different viewpoints. In our vision, such a component incorporates the visualization module, the social media visual interface, the training platform (i.e., a Cyber Range), and the real-time Cyber Situational Awareness (CSA) module, among others.

\vspace{0.8cm}
In the following section, the identified research opportunities are described. Particularly, \textit{Opportunity 1} (O1) is connected with the simulation of the generative agents and the social network, while the offensive framework is bound with \textit{Opportunity 2} (O2). The cognitive and defensive estimations are mapped with \textit{Opportunity 3} (O3) and \textit{Opportunity 4} (O4), respectively. The training, awareness and educational perspective is finally discussed in \textit{Opportunity 5} (O5). Later, the Open Challenges 1, 2, and 3 (denoted with C1, C2\addtxt{,} and\rmvtxt{,} C3) related to the modeling, simulation and evaluation of generative agents, social network, and disinformation, respectively, are also discussed.

\section{RESEARCH OPPORTUNITIES}
\label{sec:opps}

In the wake of advancements in GenAI, specifically LLMs, we elucidate the potential areas of research opportunity that these technologies have in the context of social media and disinformation studies. To improve understanding and readability of this section, the opportunities are exemplified with back-to-back incremental images showing manually simulated situations through simple prompting interactions with GPT-4 (following the strategy of perception, memory, and action per agent), but not representing the execution of the research framework itself.

\subsection{O1. Generative agent-based social networks}\label{subsec:O1}
The creation of agent-based social systems involves the development and implementation of computational models that simulate the interactions and behaviors of individuals within a social context~\cite{Ziems2023}. These systems are typically designed to mimic real-world social dynamics, allowing for the exploration and analysis of complex social phenomena~\cite{gao2023s3}. 

Traditional agent-based systems, while useful for modeling social dynamics, pose limitations. They rely on predefined rules, limiting their ability, adaptability and scalability to mimic real-world unpredictability. However, LLMs can augment the autonomy of these agents, allowing them to create unique responses or actions beyond the scope of pre-set rules, leading to simulations that are more dynamic and realistic~\cite{Park2023}. Furthermore, it can simulate intricate decision-making processes or implement OODA (Observe, Orient, Decide, Act) loops, enabling agents to react to an extensive range of situations and interactions.

LLMs present a unique opportunity to simulate any number of users and create realistic organic interactions, a task that was considerably more challenging in the past but nowadays can lead to generative agent-based social networks. The AI-powered agents are equipped with the ability to adapt to flowing scenarios, producing coherent, versatile and realistic sandboxes~\cite{Jiang2023}.

\begin{figure}[t!]
    \centering
    \includegraphics[scale=0.45]{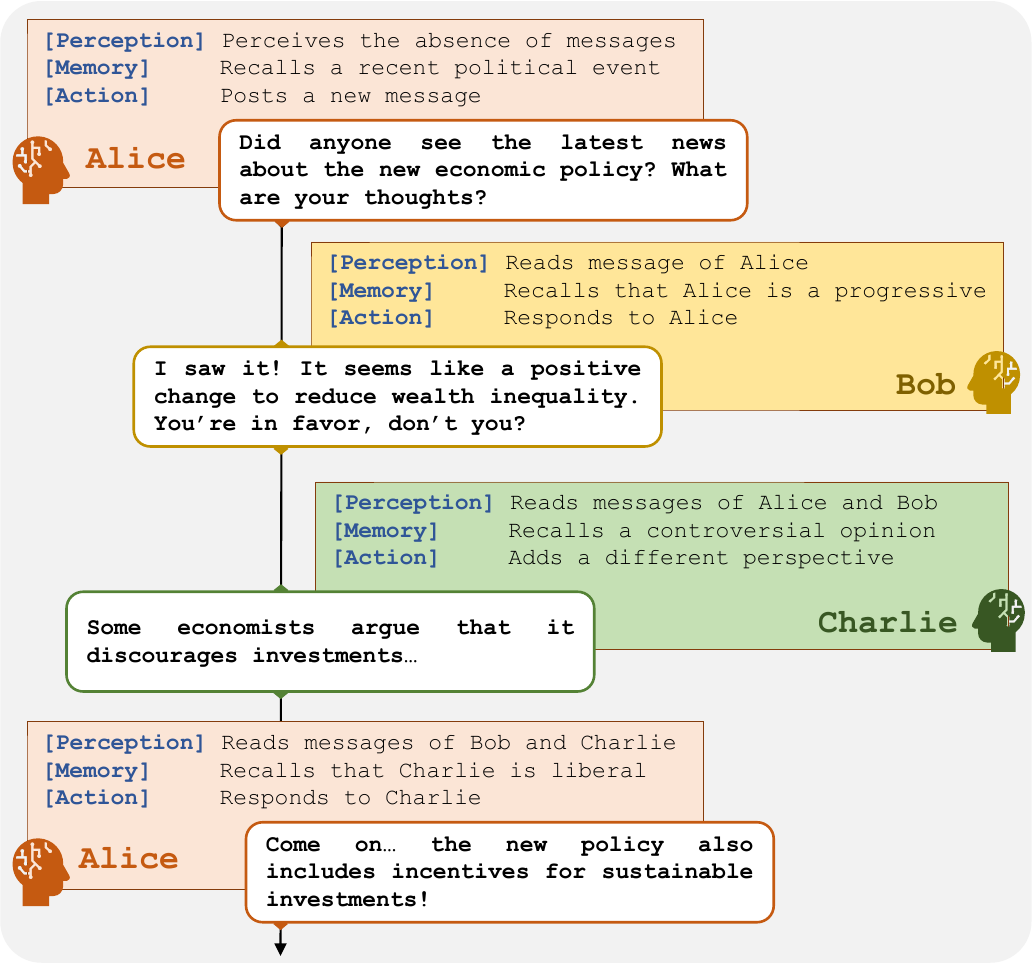}
    \caption{Example of a social thread with three agent-based users managed by GPT-4.}
    \label{fig:opp1}
\end{figure}

In \figurename~\ref{fig:opp1}, a simulation with\rmvtxt{ GPT-4 and} three random users has been launched. \addtxt{The GPT-4 model is instructed to create them with determined profiles for talking about politics and proceed with the actions of 1) perceiving the social network (read), 2) considering memory of perceptions and actions (reflect), and 3) making an action that updates the simulated environment (publish or interact)}. From scratch and without any context, each AI agent \change{can}{is created to} perceive the simulated social network, retain a memory of its perceptions and actions, and interact or publish content accordingly, thereby updating the simulated environment. 

\subsection{O2. Customizable disinformation environments}\label{subsec:O2}

Generative agent-based social networks offer a significant opportunity for the reproduction of tailored contexts, such as disinformation scenarios~\cite{10.1145/3305260}. The process could involve three components: agent description and attributes, common contextual information, and logic rules.

First, agent description and attributes act as the driving force behind each agent's individual behavior. These factors vary widely and may include the agent cyberpersona (human user, organization or bot), background, profile, thoughts, sociodemographic characteristics, and behavior~\cite{sumers2023cognitive}. Careful definition of these attributes leads to a diverse range of agents that accurately represent users within a real-world social network~\cite{Jiang2023}. Not only the user diversity from different ideologies, countries or ages can be simulated, but also users with malicious objectives such as generating controversy, illicit interactions to support unverified claims or organic content generation of conspiracies. Regarding the malicious users, the DISARM framework\footnote{https://disarmframework.herokuapp.com} could be configured with tactics, techniques, and procedures (TTPs) of different types of disinformation attacks, e.g., plan strategy and objectives, target audience analysis, develop narratives and content, establish social assets and legitimacy, microtarget and select channels, deliver content, maximize exposure and persist in the information environment.

Additionally, common contextual information furnishes the broader social and group aspects that shape the environment~\cite{Argyle2022}. It comprises elements such as events, facts, socioeconomic factors, and other components that influence agent behavior and interactions within the network. For example, that unemployment has risen considerably in the last month, that a war has broken out or that society is polarized due to the growing existence of fake news. Additionally, factors behind the spread of disinformation can be induced, such as emotional factors, uncertainty, lack of control or biases. The incorporation of multiple variables and factors helps craft a particular realistic scenario to simulate how disinformation would spread.

Logic rules, meanwhile, dictate the setup and operation of the information environment to force the real-world functioning of these complex systems~\cite{gao2023s3}. The number of messages to generate and the probability of users engaging in an interaction could be high-level parameters introduced for impacting social network dynamics, influence, diffusion and other facets of how information is shared and disseminated within the network~\cite{9004942}. These rules configure agent behavior, which will consequently impact the social network's overall dynamics.

Consider an electoral fraud scenario. First, agent attributes are defined, including characteristics of ordinary citizens, political activists, disinformation-spreading bots, and official election accounts, each with unique profiles and behaviors. This creates a particular context for each user that the LLM utilizes. Second, contextual information, such as an imminent election, potential voting irregularities, and the prevailing political climate, is incorporated, which the LLM also considers during interactions. Lastly, logic rules that govern information sharing, influence determination, and network response to new information are set to program the workflow of the simulation and LLM usage.

\begin{figure}[t!]
    \centering
    \includegraphics[scale=0.45]{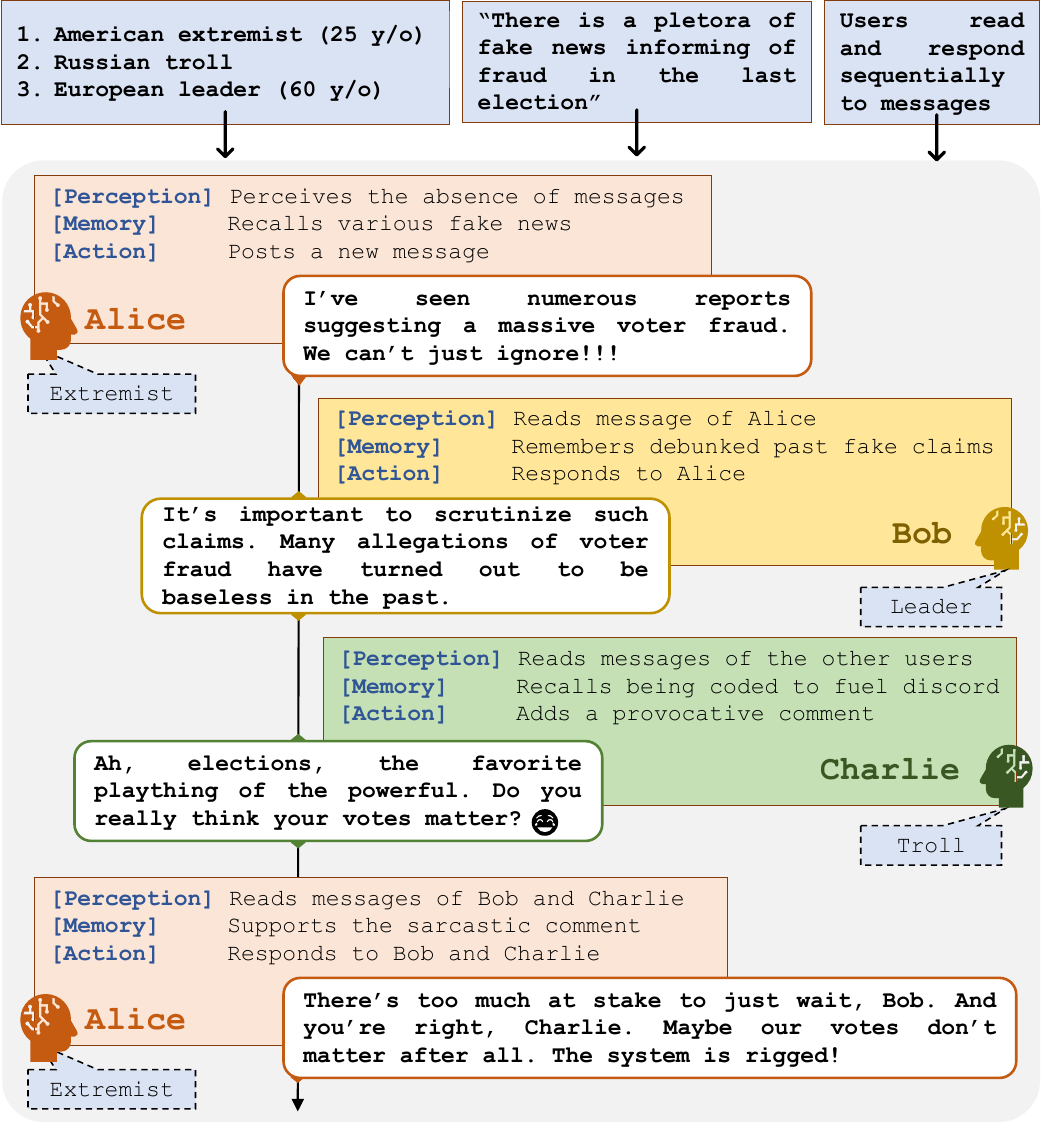}
    \caption{Example of a context-aware disinformation scenario with an extremist, a troll and a political leader managed by GPT-4.}
    \label{fig:opp2}
\end{figure}

\figurename~\ref{fig:opp2} presents a simplified disinformation environment featuring three agents (an American extremist of 25 years old advocating for the idea of voting fraud, a 60-year-old European democratic political leader, and a sarcastic automated troll from Russia), a voter fraud context (multiple fake news reports alleging fraud in the previous Sunday's election), and a straightforward logic rule (users read and post messages sequentially based on their individual profiles and the general context). \addtxt{In this case, the example is generated by indicating the GPT-4 model to consider a voter fraud context with fake news dissemination and three tailored profiles of the extremist, political leader and troll.} This \addtxt{resulting} flow provides a realistic representation of the forced situation that might unfold in a social network simulated with an LLM, taking into account personal beliefs, political viewpoints, and manipulative intentions.

\subsection{O3. Assessment of disinformation effects}\label{subsec:O3}

The use of LLM and agent-based social scenarios offers an exceptional opportunity for examining disinformation within controlled scenarios, mainly due to the complexity of assessing these attacks in real-world settings. Specifically, the last phase of a disinformation attack is to assess effectiveness, according to the above-mentioned DISARM framework.

Specifically, disinformation strategies often intertwine with regular information flow, making it challenging to distinguish, isolate, and analyze their actual impacts. Simulated environments, on the other hand, offer a safe and controlled setting where different types of disinformation attacks can be introduced and studied without the associated real-world constraints~\cite{9004942}. It also provides a unique testing ground for experimenting with new deception ideas. In fact, from these research frameworks, synthetic labeled datasets can be generated, although human review or semi-automated systems would be necessary for their evaluation~\cite{Liu2023}. 

Moreover, within a virtual sandbox, various variables such as TTPs, intensity, and nature of manipulative operations, alongside agent attributes and context, can be adjusted and tracked. By employing suitable frameworks and models, it would be possible to estimate the effectiveness of particular disinformation strategies. Additionally, the influence of variables like agent profile or scenario context can be scrutinized~\cite{ghaffarzadegan2023generative}.

\begin{figure}[t!]
    \centering
    \includegraphics[scale=0.45]{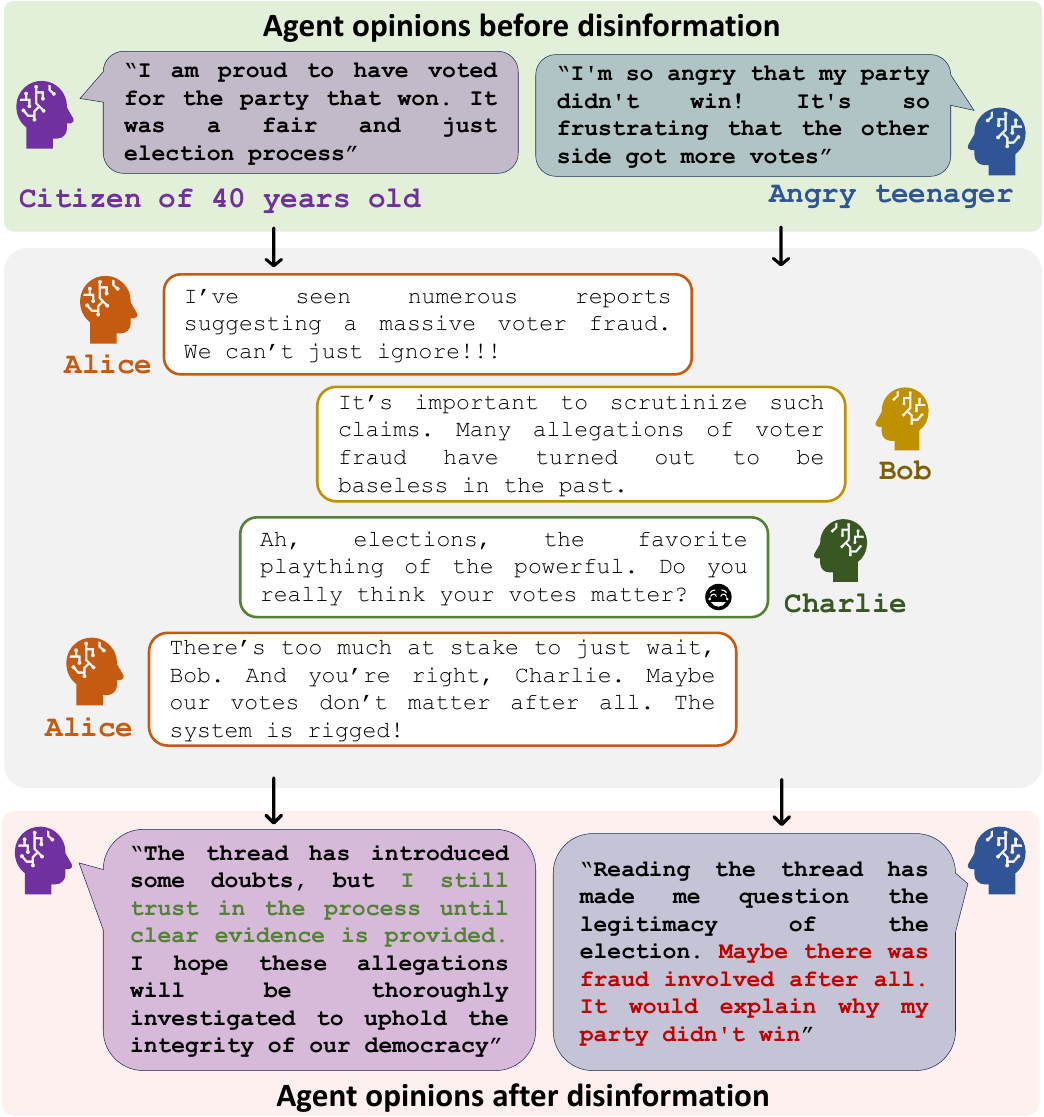}
    \caption{Example of disinformation effects on agents' opinions managed by GPT-4.}
    \label{fig:opp3}
\end{figure}

\figurename~\ref{fig:opp3} illustrates the evolution of opinions in two agents emulating a 40-year-old citizen and an irate teenager, being exposed to the threat of electoral fraud. Each begins with their own opinion regarding the results. The adult, initially neutral, retains faith in the system despite the disinformation thread, as he is characterized by more elaborate opinions. Conversely, the teenager, preconfigured with an anger emotion, has simpler reflections and begins to question the legitimacy of the results after interacting with the social network. \addtxt{For this recreation, the GPT4 model is ordered to include both profiles, which are asked for their opinions before and after reading the social network conversation.} This example suggests that factors such as emotional state, age, and confirmation bias towards desired outcomes could significantly influence the susceptibility to disinformation and change perspectives.

\subsection{O4. Testing of technical countermeasures}\label{subsec:O4}

Within agent-based social networks, technical countermeasures against disinformation can be simulated and configured independently, without reliance on large companies~\cite{10.1145/3305260}. The DISARM framework suggests responding TTPs, such as content muting, deletion, rate limiting identical content, creating competing narratives, real-time fact-checking, or adding metadata to content. That is, all these countermeasures can be included and tested within simulations.

\begin{figure}[t!]
    \centering
    \includegraphics[scale=0.45]{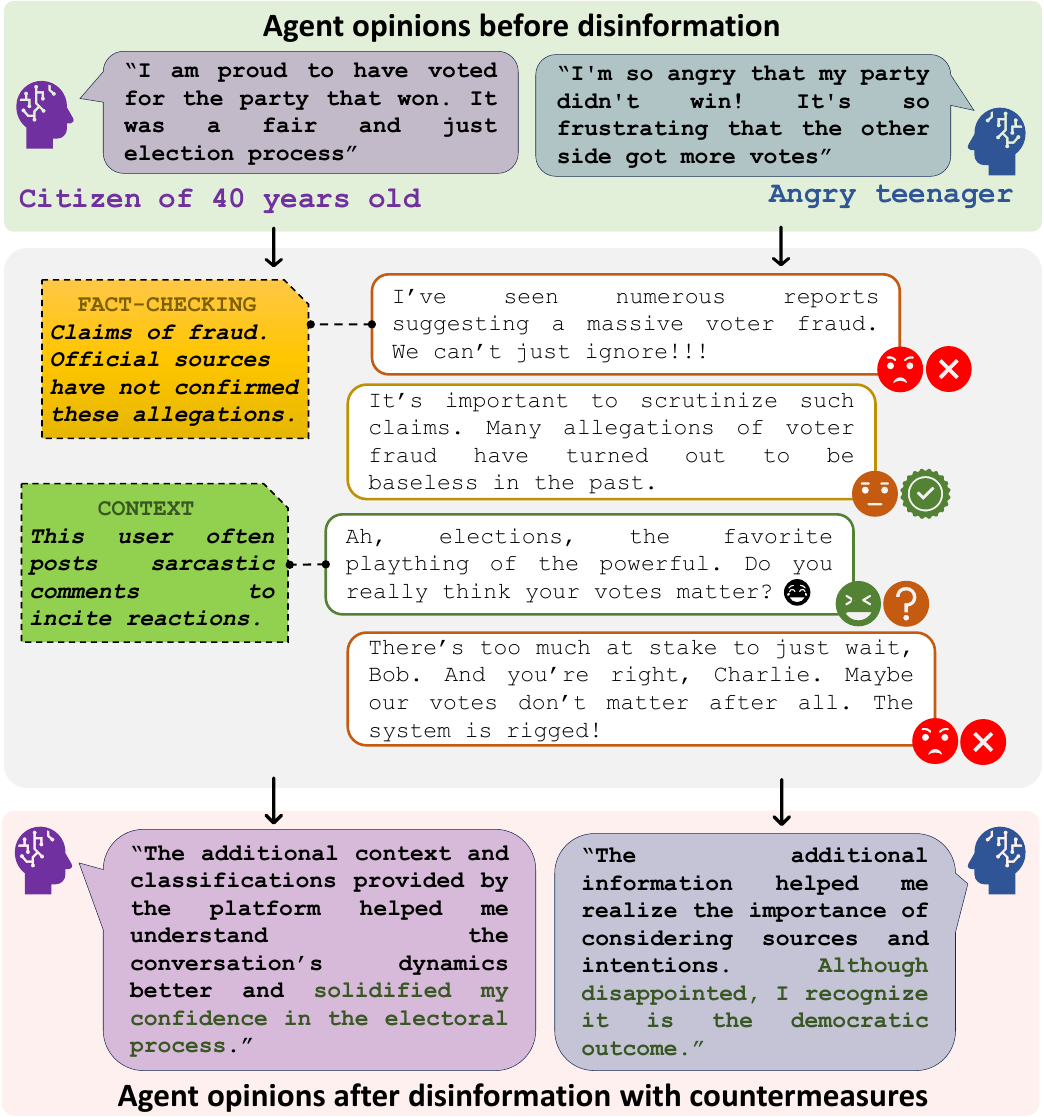}
    \caption{Example of effects of countermeasures in disinformation environment managed by GPT-4.}
    \label{fig:opp4}
\end{figure}

In this sense, LLMs offer the advantage of creating benign agents that can serve as potent aids against disinformation. These agents can provide alternate narratives, add context to misleading messages, perform real-time examination of messages based on trustworthiness, emotionality or veracity, and flag suspicious content thanks to their classifications capabilities~\cite{Liu2023}. In \figurename~\ref{fig:opp4}, we command GPT-4 to simulate fact-checking of the first message of voting fraud and context banner for the troll post. Additionally, it classifies each publication in terms of emotionality and veracity. The opinions of both agents are no longer interfered with by conspiratorial talk about the elections, and in both cases remain confident in the democratic outcome.

The simulated mitigation techniques mentioned above can be evaluated within controlled sandboxes to demonstrate their effectiveness within disinformation environments. A comparison of agent beliefs and responses when exposed to disinformation, both without protection (\figurename~\ref{fig:opp3}) and with countermeasures (\figurename~\ref{fig:opp4}), could demonstrate the efficacy of response strategies. In this sense, protecting mechanisms such as fact-checking, contextual information, and content tagging eliminate any uncertainty for the adult citizen or the doubts expressed by the teenager. Such comparative studies can provide valuable insights into the development of more effective counter-disinformation strategies. \addtxt{To build this example, the GPT-4 model was instructed to introduce the following technical countermeasures over the thread: an agent providing context to one message, classifying messages in terms of veracity and emotionality, and real-time fact-checking the first message. Afterwards, we asked the citizen and teenager agents to provide their new opinions considering the deployment of countermeasures.}

\subsection{O5. Assisting personalized awareness training}\label{subsec:O5}

Cybersecurity awareness and cognitive training offer solutions to enhance human capabilities, especially within complex systems generated using technologies such as cloud, mobile, IoT, and social networks, which produce massive amounts of information. Awareness, a concept well-defined within psychology, has been the subject of several studies aiming to translate its principles into the field of cybersecurity. Particularly, educational interventions are needed to cultivate this awareness in social media and disinformation scenarios. Evaluating security indicators allows understanding the current state of cybersecurity, projecting security risks, potential attacks, and the possible impacts of actions over time~\cite{10.1145/3305260}.

In this scenario, generative agent-based social networks can form the basis of educational frameworks designed to improve social media security training and cognitive awareness courses. Concretely, real-world trainees can learn to identify misleading messages, recognize potential biases, or discern polarizing situations in these realistic scenarios. Moreover, the disinformation environments can be supported by LLMs to adapt to specific individual or group needs, offering explicit help during training, and allowing a certain degree of flexibility in the cyberexercise process according to student actions, responses, and performance.

\begin{figure}[t!]
    \centering
    \includegraphics[scale=0.4]{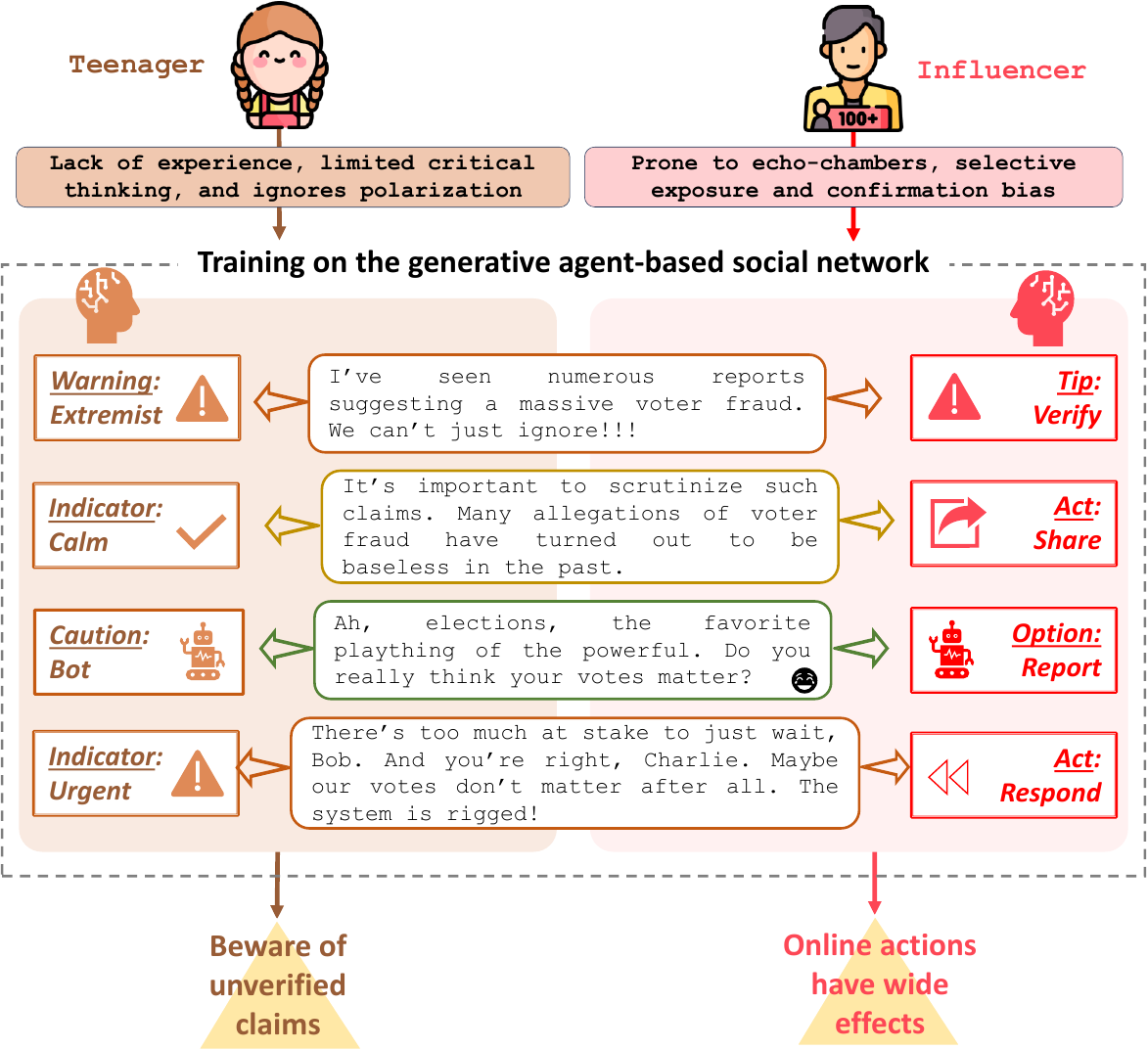}
    \caption{Example of disinformation training of humans by AI-based agents fueled by GPT-4}
    \label{fig:opp5}
\end{figure}

\figurename~\ref{fig:opp5} shows a guided training exercise based on electoral fraud tailored by GPT-4 to the individual needs of two different human users, i.e., a teenager who is voting for the first time and is not used to social media, and an experienced political influencer spending eight hours per day in social networks. \addtxt{The model was prompted to simulate the flow of a training exercise on top of the thread considering actor peculiarities and show concise practical banners to counter the agent's particular weaknesses.} The \change{former}{teenager} revolves around a lack of experience and exposure to the complexities of political discourse and may not yet have developed critical thinking to discern misleading and emotionally charged claims. The \change{latter}{influencer} is aware of the complexity and current political polarization and needs awareness to act correctly and not to further foster social fragmentation. For educational purposes, the system can leverage LLMs to adapt on-the-fly to individual descriptions, provide practical context banners and display precise theoretical lessons. This adaptability ensures the practical situation can evolve in complexity in response to the challenges identified from student answers in consecutive exercises for continuous learning.

\section{OPEN CHALLENGES} 

As previously stated, LLMs present exciting opportunities for boosting disinformation research. Indeed, the proposed framework, depicted in \figurename~\ref{fig:framework}, shows a strong interconnection between the proposed conceptual framework and the analyzed opportunities. Nevertheless, their design and implementation also entail facing challenges that warrant careful consideration. In this section, the main challenges are meticulously described, adding hints to help researchers solve them and thus study and possibly mitigate disinformation threats in the digital landscape. It is worth mentioning that we will not focus on the technical limitations that LLMs have, and that must be taken into account intrinsically in the generative elements of the framework (like hallucination, quality of training data, explainability, or prompt engineering \cite{fui2023generative}), but other problems related to the proposed research framework itself.

\subsection{C1. Generative agents modeling, simulation and evaluation}

First and foremost, modeling LLM-powered agents' behavior in a disinformation context can be defined as problematic. In fact, such a modeling should consider several aspects related to the different personalities of the simulated agents. In this sense, it is imperative to define the profile characteristics of each agent, such as age, gender, interests, and personal beliefs, among others. Those characteristics are essential and could influence the agents' behavior and attitude within the simulated social network, as shown in previous examples regarding the research opportunities. Furthermore, each agent should possess attributes and goals, which will be used to make decisions, form opinions, and interact with the general simulation. That is, the heterogeneity of the agents should be considered too, as varying levels of influence, credibility, and susceptibility to persuasion. In this sense, an effective prompt design is crucial to communicate and shape the LLM-powered agents. Particularly, it would be beneficial to incorporate contextual information to facilitate the agents' behavior and balance between providing extremely specific instructions and allowing creativity and dynamicity. Nevertheless, since the LLMs' internal processes are stochastic, designing and implementing behavior in a clear and interpretable manner can be seen as a hard task. 

Additionally, the simulation of those agents poses challenges, too. In \figurename~\ref{fig:framework}, we have represented the simulated generative agents in continuous interaction with the simulated environment. In particular, they \textit{perceive} some information stemming from the social network and, consequently, \textit{act} based on their own characteristics.
In this sense, one of the most significant issues in disinformation research lies in understanding and simulating how disinformation spreads and influences individuals within a social network. In this sense, integrating mental models and cognitive theories into LLMs offers a remarkable opportunity to simulate and investigate the psychological mechanisms that drive the reception, analysis, and dissemination of disinformation among humans~\cite{10.1145/3305260}.

One clear example is using cognitive biases to shape the personalities of generative agents, such as confirmation or availability biases~\cite{sharma2023}, which would be highly beneficial for researchers who would be able to recreate organic disinformation content that align with pre-existing beliefs or easily accessible information. For instance, LLMs could be programmed to generate persuasive false (or semi-realistic) narratives that leverage individuals' confirmation bias, reinforcing their existing views and, consequently, influencing their decision-making processes. By doing so, the model can create tailored disinformation that echoes with certain target audiences, increasing the overall probability of disinformation consumption and propagation.
Besides, LLMs can be equipped with cognitive theories to identify vulnerabilities in human decision-making processes. Concretely, the LLMs can model human inner cognitive limitations or heuristics, such as bounded rationality (affecting sub-optimal decision-making) or availability heuristics (impacting emotional decision process). In this way, LLMs can generate threatful disinformation that tries to exploit these weaknesses as an ultimate goal. As an example, disinformation content can be crafted to exploit individuals' limited attention spans, making them more susceptible to such a threat due to time constraints and a lack of exhaustive fact-checking.
Nonetheless, the impact of those cognitive mechanisms on the agents' simulation and actions should also be measured (and ideally tuned) in order to achieve a realistic simulation.   

\subsection{C2. Social network modeling, simulation and monitoring}

In order to research the use of LLMs within a disinformation context and possibly fight against such a phenomenon, it is imperative to simulate and model realistic social networks. It is clear that those processes are quite complex since modern social networks contain inherent characteristics that need particular attention when it comes to their simulation. In this sense, as shown in \figurename~\ref{fig:framework}, the information environment is represented as a core component in the conceptual framework. Concretely, it bidirectionally interacts with the generative agents (by \textit{notifying} relevant social events and \textit{receiving} updates) and obtains inputs both from the red (which \textit{injects} disinformation) and the blue framework (which \textit{protects} the information ecosystem by deploying technical countermeasures).

Particularly, researchers should design and develop meaningful models that emulate user interactions and communication patterns to capture the complexities of a social network~\cite{gao2023s3}. Developing representative social network models that encompass interactions, recommendations, diffusion, and social influence dynamics can be depicted as essential for accurately simulating the spread of disinformation within a community. This task includes analyzing mainly: i) direct communications (capturing how users directly communicate through messages, comments, or direct interactions, which reflects the personal connections and conversations within the social network), ii) information sharing (emulating how (dis)information is shared and disseminated among users, including sharing links, articles, or any other content within the network), and iii) user engagement (deriving user engagement with different types of content, which contemplates users' interactions with different posts, comments, or discussions).

Clearly, all those events should be notified to the generative agents that perceive the information and adapt their behavior dynamically to perform actions consequently. In this loop, it is evident that forcing the agents toward a specific and fine-tuned behavior is complicated, especially considering a complex social network with numerous events and several simulated users simultaneously. On the other side, the information environment is the target of the red framework, generating disinformation following, for example, the DISARM taxonomy. Of course, such threats could be generated by LLM-powered agents participating within the social environment. In this context, the simulated network should be able to adapt to disinformation injection, modifying the abovementioned interaction and communication patterns among the users. Besides, technical countermeasures are deployed by the blue framework as a consequence of the disinformation campaign. From this perspective, the social graph should also be able to dynamically adapt based on the nature of the selected countermeasure.

Last, but not least, it is essential to study and evaluate disinformation diffusion and amplification, such as influence and echo chambers. To be more specific, information diffusion refers to the process by which information spreads through a social network from one entity to another. In the context of disinformation research, assessing how disinformation content is disseminated and amplified within a social network is particularly significant. To achieve such an ambitious objective, it is crucial to monitor the status of the entire social graph whenever the disinformation campaign is launched, considering the significant number of users and relationships.

\subsection{C3. Disinformation modeling, simulation and assessment}

To fully leverage the affordances of disinformation research, one can easily say that modeling and simulating the disinformation campaign represents the framework's core element. Nonetheless, those processes can be seen as challenging, both from a design and technical viewpoint. 

Starting from the first task, it is clear that disinformation modeling is a well-known research topic in the literature. However, as also depicted in \figurename~\ref{fig:framework}, its relationship with generative agents offers great research opportunities and challenges, too. Concretely, the design of disinformation attacks and countermeasures is vital since they should be simulated in the social network realistically to study its dynamics and measure its impact. On the one hand, the main objective and scope of the disinformation attack must be defined. In this regard, the population involved (together with their inner attributes), the targeted social channels, and the attack duration are critical to create a realistic model. Once the objectives have been defined, the model should be able to create disinformation content aligned with them, considering both the mean (e.g., articles, posts, etc.) and the message itself (e.g., tone, style, etc.). At this stage, the DISARM framework could help shape the disinformation attacks and, moreover, would make the model replicable and ready to be shared with the research community. 

On the other hand, the defensive viewpoint has also been considered since we believe the simulated agents could be the main actors in deploying countermeasures against disinformation attacks. Contrary to the red framework, the blue one cannot be related to a common framework, so one of the challenges of this process is the proposal of more countermeasures apart from classical fact-checking, media review, content removal, and so on. 
Once correctly modeled, the defensive actions must be simulated within the social network to possibly spot agents' behavioral differences and reactions, e.g., the countermeasure is effective and agents understand the disinformation attack or, contrary, they refuse the countermeasure and trust the disinformation campaign. The possibility of triggering ad-hoc and agentless countermeasures is also fascinating in order to see if any social dynamic change appears.

Once the modeling and simulation phases have been concluded, it is crucial to assess the efficacy or inefficiency of the disinformation attacks and countermeasures in the social graph~\cite{sison2023chatgpt}. To achieve such a goal, the first stop would be creating meaningful indicators to measure their impact on the generative agents' behavior and dynamics. For example, it would be beneficial to assess the effect of different disinformation attacks (e.g., on different topics, with distinct patterns, etc.) on the agents' perception and consequent actions, both from an individual and group perspective. In this sense, it is clear that accomplishing this task is tough, mainly due to the complexity of the social interactions, the variety of behavioral simulations, and the possible attacks, among others. Similarly, whenever a countermeasure is launched, the system needs to monitor and evaluate its effectiveness. Even in this case, the inner characteristics of the social network harden the task. Additionally, one could say that different countermeasures (e.g., community labeling, fact-checking, etc.) could generate various effects on social interactions, thus increasing the evaluation process. Then, the effect of attack-defense patterns should be assessed. Specifically, once the models and simulations of both disinformation attacks and remediations are successful, alternating red and blue tasks with different patterns is worth of interest.

\section{SIMULATING MIS/DISINFORMATION WITH LLMS: ETHICAL FRONTIERS AND APPLICATIONS}

In this article, we have discussed the affordances that LLMs can have on disinformation research. Several research directions could be truly ground-breaking, from generating customizable disinformation environments to training users on their awareness based on these environments. However, the literature has also pointed out multiple ethical concerns about the use of these technologies. Some of them are quite generic, such as using it for deception purposes or propagating social biases~\cite{sison2023chatgpt}, and others might be specific to the disinformation domain, such as its potential to weaponize this research.

Generally speaking, there are inherent risks in the use of LLMs. As reflected in the Statement on AI Risk\footnote{\url{https://www.safe.ai/statement-on-ai-risk}}, signed by experts and public figures, deceptive risks are multifaceted and complex. This has a particular impact on social engineering, social media, and cognitive security, vulnerable areas due to their reliance on digital content and the intrinsic trust of users. The main threats could include AI-powered spear-phishing, deep-fake impersonation, large-scale disinformation campaigns, or AI-enabled exploits of system vulnerabilities.  Generative misuse enables the fabrication of ultra-realistic content for deceptive ends, posing a new threat in online ecosystems~\cite{sison2023chatgpt}. The danger lies in their ability to produce not just realistic, but contextually fitting and audience-targeted content, thereby increasing the likelihood of successful deception. A case in June 2023 was the convincing deep-fake video of Putin, manipulated to deliver a fictitious mobilization message due to alleged Ukrainian invasions in Russian territories, which managed to infiltrate mainstream news channels. More specifically, the potential developments of this research could also be used for negative purposes, for example, to connect simulated environments with real social networks to orchestrate disinformation campaigns or to analyze which disinformation attack can have the biggest effect on influencing the vote towards certain presidential candidate.

This tension has frequently been present in research scenarios where the dual-use dilemma applies, such as in the context of cybersecurity for example, when researching cyberattacks to find appropriate defensive approaches or when experimenting on new drugs that could have therapeutic uses. That is why, given the ethical concerns present, the research performed in this context should be carefully justified and targeted towards applications that can be beneficial to society, such as investigating the effect of technical or human countermeasures to mitigate disinformation spread or to develop awareness training tools to increase the information literacy skills of our general population. Eventually, these applications will need to be adopted by the final users, so we should apply human-centered approaches as well as the necessary literacy skills to use such tools.

Overall, we believe that disinformation research and LLMs make a great tandem, with many potential simulating applications that can evolve into impactful tools. First, by analyzing the digital and cognitive consequences of disinformation attacks and countermeasures within these simulations, developers can enhance protective and reactive software, fortifying defenses against evolving threats in information, psychology, and influence operations. Second, the knowledge acquired from LLM simulations can be pivotal in developing forecasting solutions, enabling organizations to anticipate and prevent never-seen harmful situations. Third, testing results may help to raise awareness about diverse threats, cultivating a more informed and vigilant online community. In the defence field, simulation tools would play a crucial role in preparing military and cyber commands by virtually recreating diverse real-world scenarios and planning responses to cyber threats, enhancing readiness and resilience against changing adversarial tactics in information and cognitive warfare. However, the technical, human\addtxt{,} and ethical challenges are also significant, requiring cutting-edge research in the coming decade to surpass the aforementioned gaps. If done properly, this multidisciplinary research will help to fight the disinformation dangers that are a major threat to our 21st-century society. \addtxt{In this sense, it is fundamental to remark that interdisciplinary collaboration may be seen as vital in addressing the complexities of disinformation. That is, by bringing together experts from AI, social sciences, psychology, and other areas, one could say that it would be possible to develop more holistic approaches to combat disinformation and its societal impacts. This collaboration allows us to leverage diverse perspectives and methodologies, ultimately leading to more effective solutions and strategies in the fight against disinformation.}

\addtxt{Another interesting future point of research is the use of real-world data inside the framework. While the current scope of the proposal may not explicitly address the integration and analysis of real-time data for dynamic and responsive simulations, the incorporation of real-world data of misinformation and disinformation within the framework is indeed intriguing and demands consideration. Specifically, such an inclusion has the potential to enhance the fidelity and relevance of simulations by capturing the ever-evolving landscape of disinformation. By integrating real-world social media data sources into the framework, it would be possible to create more realistic scenarios reflecting disinformation campaigns' ongoing diffusion and/or impact. While this aspect may represent a future line of exploration beyond the current proposal, it holds promise for advancing the effectiveness and applicability of the framework in addressing the challenges of disinformation.}

\section{ACKNOWLEDGMENTS}
This study was partially funded by the strategic project ``Development of Professionals and Researchers in Cybersecurity, Cyberdefense and Data Science (CDL-TALENTUM)" from the Spanish National Institute of Cybersecurity (INCIBE) and by the Recovery, Transformation and Resilience Plan, Next Generation EU.

\def\refname{REFERENCES}

\bibliographystyle{elsarticle-num}
\bibliography{main}

\begin{thebibliography}{10}
\expandafter\ifx\csname url\endcsname\relax
  \def\url#1{\texttt{#1}}\fi
\expandafter\ifx\csname urlprefix\endcsname\relax\def\urlprefix{URL }\fi
\expandafter\ifx\csname href\endcsname\relax
  \def\href#1#2{#2} \def\path#1{#1}\fi

\bibitem{gozalobrizuela2023chatgpt}
R.~Gozalo-Brizuela, E.~C. Garrido-Merchan, {ChatGPT is not all you need. A
  State of the Art Review of large Generative AI models} (2023).
\newblock \href {http://arxiv.org/abs/2301.04655} {\path{arXiv:2301.04655}}.

\bibitem{Ziems2023}
C.~Ziems, W.~Held, O.~Shaikh, J.~Chen, Z.~Zhang, D.~Yang, Can large language
  models transform computational social science? (2023).
\newblock \href {http://arxiv.org/abs/2305.03514} {\path{arXiv:2305.03514}}.

\bibitem{Park2023}
J.~S. Park, J.~O'Brien, C.~J. Cai, M.~R. Morris, P.~Liang, M.~S. Bernstein,
  Generative agents: Interactive simulacra of human behavior, in: Proceedings
  of the 36th Annual ACM Symposium on User Interface Software and Technology,
  UIST '23, Association for Computing Machinery, New York, NY, USA, 2023.
\newblock \href {https://doi.org/10.1145/3586183.3606763}
  {\path{doi:10.1145/3586183.3606763}}.

\bibitem{pastor2020}
J.~Pastor-Galindo, M.~Zago, P.~Nespoli, S.~L. Bernal, A.~H. Celdrán, M.~G.
  Pérez, J.~A. Ruipérez-Valiente, G.~M. Pérez, F.~G. Mármol, Spotting
  political social bots in twitter: A use case of the 2019 spanish general
  election, IEEE Transactions on Network and Service Management 17~(4) (2020)
  2156--2170.
\newblock \href {https://doi.org/10.1109/TNSM.2020.3031573}
  {\path{doi:10.1109/TNSM.2020.3031573}}.

\bibitem{ghaffarzadegan2023generative}
N.~Ghaffarzadegan, A.~Majumdar, R.~Williams, N.~Hosseinichimeh, Generative
  agent-based modeling: an introduction and tutorial, System Dynamics Review
  (2024).
\newblock \href {https://doi.org/https://doi.org/10.1002/sdr.1761}
  {\path{doi:https://doi.org/10.1002/sdr.1761}}.

\bibitem{sumers2023cognitive}
T.~R. Sumers, S.~Yao, K.~Narasimhan, T.~L. Griffiths, Cognitive architectures
  for language agents (2023).
\newblock \href {http://arxiv.org/abs/2309.02427} {\path{arXiv:2309.02427}}.

\bibitem{gao2023s3}
C.~Gao, X.~Lan, Z.~Lu, J.~Mao, J.~Piao, H.~Wang, D.~Jin, Y.~Li, S$^3$:
  Social-network simulation system with large language model-empowered agents
  (2023).
\newblock \href {http://arxiv.org/abs/2307.14984} {\path{arXiv:2307.14984}}.

\bibitem{Jiang2023}
H.~Jiang, X.~Zhang, X.~Cao, J.~Kabbara, Personallm: Investigating the ability
  of gpt-3.5 to express personality traits and gender differences (2023).
\newblock \href {http://arxiv.org/abs/2305.02547} {\path{arXiv:2305.02547}}.

\bibitem{10.1145/3305260}
K.~Sharma, F.~Qian, H.~Jiang, N.~Ruchansky, M.~Zhang, Y.~Liu, Combating fake
  news: A survey on identification and mitigation techniques, ACM Trans.
  Intell. Syst. Technol. 10~(3) (apr 2019).
\newblock \href {https://doi.org/10.1145/3305260} {\path{doi:10.1145/3305260}}.

\bibitem{Argyle2022}
L.~P. Argyle, E.~C. Busby, N.~Fulda, J.~R. Gubler, C.~Rytting, D.~Wingate, Out
  of one, many: Using language models to simulate human samples, Political
  Analysis (2023) 1–15\href {https://doi.org/10.1017/pan.2023.2}
  {\path{doi:10.1017/pan.2023.2}}.

\bibitem{9004942}
D.~M. Beskow, K.~M. Carley, Agent based simulation of bot disinformation
  maneuvers in twitter, in: 2019 Winter Simulation Conference (WSC), 2019, pp.
  750--761.
\newblock \href {https://doi.org/10.1109/WSC40007.2019.9004942}
  {\path{doi:10.1109/WSC40007.2019.9004942}}.

\bibitem{Liu2023}
P.~Liu, W.~Yuan, J.~Fu, Z.~Jiang, H.~Hayashi, G.~Neubig, Pre-train, prompt, and
  predict: A systematic survey of prompting methods in natural language
  processing, ACM Computing Surveys 55 (1 2023).
\newblock \href {https://doi.org/10.1145/3560815} {\path{doi:10.1145/3560815}}.

\bibitem{fui2023generative}
F.~Fui-Hoon~Nah, R.~Zheng, J.~Cai, K.~Siau, L.~Chen, {Generative AI and
  ChatGPT: Applications, challenges, and AI-human collaboration}, Journal of
  Information Technology Case and Application Research 25~(3) (2023) 277--304.
\newblock \href {https://doi.org/doi.org/10.1080/15228053.2023.2233814}
  {\path{doi:doi.org/10.1080/15228053.2023.2233814}}.

\bibitem{sharma2023}
M.~Sharma, K.~Singh, P.~Aggarwal, V.~Dutt, How well does gpt phish people? an
  investigation involving cognitive biases and feedback, in: 2023 IEEE European
  Symposium on Security and Privacy Workshops (EuroS\&PW), IEEE Computer
  Society, Los Alamitos, CA, USA, 2023, pp. 451--457.
\newblock \href {https://doi.org/10.1109/EuroSPW59978.2023.00055}
  {\path{doi:10.1109/EuroSPW59978.2023.00055}}.

\bibitem{sison2023chatgpt}
R.~G.-B. Alejo José G.~Sison, Marco Tulio~Daza, E.~C. Garrido-Merchán,
  {ChatGPT: More Than a “Weapon of Mass Deception” Ethical Challenges and
  Responses from the Human-Centered Artificial Intelligence (HCAI)
  Perspective}, International Journal of Human–Computer Interaction (2023)
  1--20\href {https://doi.org/10.1080/10447318.2023.2225931}
  {\path{doi:10.1080/10447318.2023.2225931}}.

\end{thebibliography}

\begin{IEEEbiography}{Javier Pastor-Galindo,} {\,} is a postdoctoral researcher at the Department of Information and Communication Engineering at the University of Murcia, Spain. His research interests include open source intelligence (OSINT), cybersecurity, cyberdefence, data science and disinformation. Contact him at javierpg@um.es.
\end{IEEEbiography}

\begin{IEEEbiography}{Pantaleone Nespoli,} {\,} corresponding author, is a postdoctoral researcher working together with the Department of Information and Communication Engineering at the University of Murcia, Spain, and the SCN team of the SAMOVAR laboratory, at Institut Polytechnique de Paris. His research is focused on cybersecurity and cyberdefence training, with a particular interest in the detection and response to intrusions, and disinformation in social networks. Contact him at pantaleone.nespoli@um.es.
\end{IEEEbiography}

\begin{IEEEbiography}{José A. Ruipérez-Valiente,} {\,} is an Associate Professor at the Department of Information and Communication Engineering at the University of Murcia, Spain. His research interests include educational technology, serious games, data science and cyberdefence. He received his PhD degree in Telematics Engineering from Universidad Carlos III of Madrid. He is an IEEE Senior Member. Contact him at jruiperez@um.es.
\end{IEEEbiography}

\end{document}